\documentclass[journal=jpclcd,manuscript=letter]{achemso}
\usepackage{xcolor}
\usepackage[normalem]{ulem} 

\definecolor{darkgreen}{RGB}{0,150,0}
\newcommand{\corr}[1]{} 
\newcommand{\newtxt}[1]{\textcolor{black}{#1}} 

\title{Simulating Attochemistry: Which Dynamics Method to Use?}
\author{Thierry Tran}
\author{Anthony Fert\'{e}}
\author{Morgane Vacher}
\email{morgane.vacher@univ-nantes.fr}
\affiliation[Unknown University]{Nantes Universit\'{e}, CNRS, CEISAM UMR 6230, F-44000 Nantes, France
\\\vspace{10pt}}

\begin{document}
\begin{tocentry}
\begin{center}
    \includegraphics{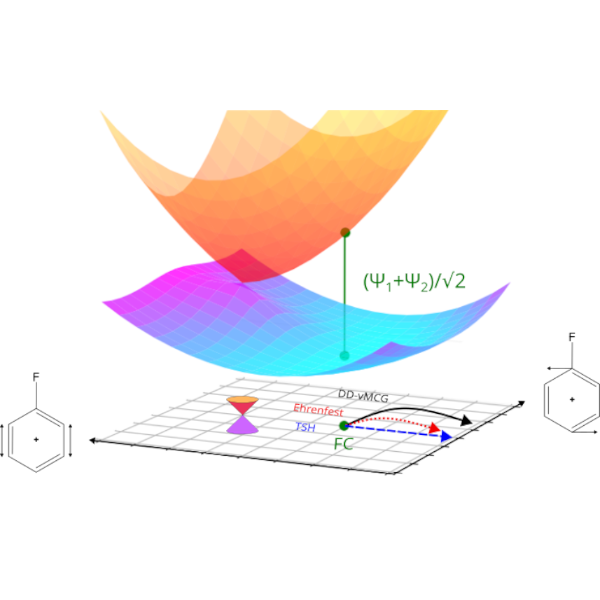}
\end{center}
\end{tocentry}
\begin{abstract}
Attochemistry aims to exploit the properties of coherent electronic wavepackets excited via attosecond pulses, to control the \corr{outcome of photochemical reactions}\newtxt{formation of photoproducts}. \corr{The simulation of the time evolution of}\newtxt{S}uch\corr{a} molecular process\newtxt{es} can in principle be \corr{performed using}\newtxt{simulated with} various nonadiabatic dynamics methods, yet the impact of the approximations underlying the methods is rarely assessed. 
\corr{Here, we use a high-accuracy DD-vMCG quantum dynamics calculation as a reference for evaluating the performances of widely used mixed quantum-classical approaches, the Tully surface hopping and classical Ehrenfest methods, using ionization \corr{and excitation}\newtxt{of fluorobenzene} to the two lowest-energy electronic states\corr{of fluorobenzene} as a test case.}
\newtxt{The performances of widely used mixed quantum-classical approaches, the Tully surface hopping, and classical Ehrenfest methods are evaluated against the high-accuracy DD-vMCG quantum dynamics. This comparison is conducted on the valence ionization of fluorobenzene.}
Analy\corr{s}\newtxt{z}ing the nuclear motion induced in the branching space of the nearby conical intersection, the results show that the mixed quantum-classical methods reproduce quantitatively the average motion of a quantum wavepacket when initiated on a single electronic state. However, they fail to properly capture the nuclear motion induced by an electronic wavepacket \corr{in the direction of}\newtxt{along} the derivative coupling, the latter originating from the quantum electronic coherence property -- key to attochemistry.
\end{abstract}


In 2001, the first attosecond ($10^{-18}$ s) domain pulses were generated.\cite{Hentschel-2001,Paul-2001,Frank2012,Duris-2020}
 Their application to polyatomic molecules \corr{has given birth to}\newtxt{sparked the emergence of} the field \corr{of}\newtxt{known as} ``attochemistry''.\cite{Remacle-2006,Calegari-2014,Kraus-2015,Barillot-2021,Mansson-2021} 
One distinctive characteristic of these ultrashort pulses is their broad energy bandwidth, capable of coherently exciting/ionizing to multiple electronic states of molecules.\cite{Breidbach-2005,Cederbaum-1999,Nisoli-2017} 
Electronic wavepackets are non-stationary states: treating only the electronic degrees of the system, the electronic density of a wavepacket oscillates with a period inversely proportional to the energy difference between the populated \newtxt{electronic eigen}states.\cite{Cederbaum-1999} \corr{By}\newtxt{In particular, due to} interference \newtxt{effect}, the electronic distribution of an electronic wavepacket is not the simple average of the individual electronic distributions of the different states.
\corr{Consequently}\newtxt{As a result}, an electronic wavepacket can exhibit properties that markedly differ from those of pure states populated individually, potentially resulting in distinct chemical reactivity.
One of the goals of attochemistry is to manipulate the outcome of photochemical reactions through the initial coherent excitation of electronic wavepackets with specific compositions.\cite{Merritt-2021,FerVac-INC-22} 
This concept, also referred to as charge-directed dynamics in the literature, has been explored experimentally and computationally in small systems.\cite{Roudnev-2004,Kling-2006,Kling-2008,Rozgonyi-2008,Lepine-2014,Valentini2020}
Using the simplest case of \corr{an}excitation to a coherent superposition of two electronic states, the ensuing nuclear dynamics manifest as two components: an intrastate component combining the two adiabatic gradients and an interstate component along the derivative coupling. Notably,\corr{it is} the second contribution, \newtxt{which} aris\newtxt{es}\corr{ing} from \corr{the} electronic coherence\corr{ and specifically the relative phase between the two states}, \corr{that} gives emergent properties to electronic wavepackets compared to traditional photochemistry.\cite{Tran-2021}

Theoretically, simulating photochemical reactions induced by electronic wavepackets implies propagating coherently nuclear wavefunctions initiated on multiple electronic states.\cite{Cederbaum-1999}
The most \corr{suited}\newtxt{accurate} family of \corr{dynamics}method\newtxt{s} to simulate such processes is fully quantum dynamics among which the Multi-Configuration Time-Dependent Hartree (MCTDH)\cite{Meyer-1990} approach is \corr{the most} commonly employed.
However, \corr{it comes with a high}\newtxt{the} computational cost \corr{, scaling}\newtxt{scales} exponentially with the size of the system on top of requiring pre-fitted potential energy surfaces (PES). This limits the theoretical studies to small molecules/model systems,\cite{Nikodem-2017,Despre-2018,Schnappinger2021,DeyKulWor-PRL-22} or medium-size molecules in reduced-dimensionality.\cite{Schuppel2020,Valentini2020} Simulations on model systems are suitable for the description of mostly rigid systems but hardly generalizable to photoreactions in polyatomic molecules involving, for instance, isomeri\corr{s}\newtxt{z}ation or dissociation. 
\corr{Besides}\newtxt{Addtionally}, a calculation on a system with reduced degrees of freedom may miss some relevant features and bias the predicted dynamics.\cite{Gomez-2019} To overcome these bottlenecks, the alternative in attochemistry has been to perform simulations with frozen nuclei (thus unable to investigate charge-directed reactivity),\cite{Cederbaum-1999} or treating all nuclear degrees of freedom with mixed quantum-classical methods.\cite{Meisner-2015,Vacher-2015-JCP,Vacher-2016-FD} 
A popular example of such methods is Tully surface hopping (TSH) where a swarm of independent trajectories \corr{are}\newtxt{is} propagated on adiabatic PES and able to hop between them.
\cite{Tully-1971,Tully1990,Merritt-2023} 
Another common method is the classical Ehrenfest\cite{Li2005} where a mean-field is employed to propagate the independent trajectories on \corr{a }single time-dependent PES corresponding to the superposition\newtxt{s} of electronic eigenstates.
\corr{The}\newtxt{While} mixed quantum-classical approaches are computationally more affordable for larger molecules\corr{; however}, they do not treat correctly the electronic coherence -- a fundamental key component for attochemistry. 
To \corr{combine the complete}\newtxt{balance a comprehensive} quantum treatment of the electrons and nuclei \corr{and retain a}\newtxt{with} reasonable computational cost, one can use an approach \corr{with}\newtxt{based on} Gaussian wavepackets (GWP) such as the Direct Dynamics variational MultiConfigurational Gaussian (DD-vMCG)\cite{Richings-2015} method: \newtxt{in that method,} both nuclei and electrons are treated quantum mechanically and the localized nature of the GWP requires \corr{only}the evaluation of the PES \newtxt{up to second order} at the center of the basis functions \newtxt{only (local harmonic approximation),} allowing an \textit{on-the-fly} quantum simulation.

\begin{figure}[h!]
    \centering
    \includegraphics[width=\textwidth]{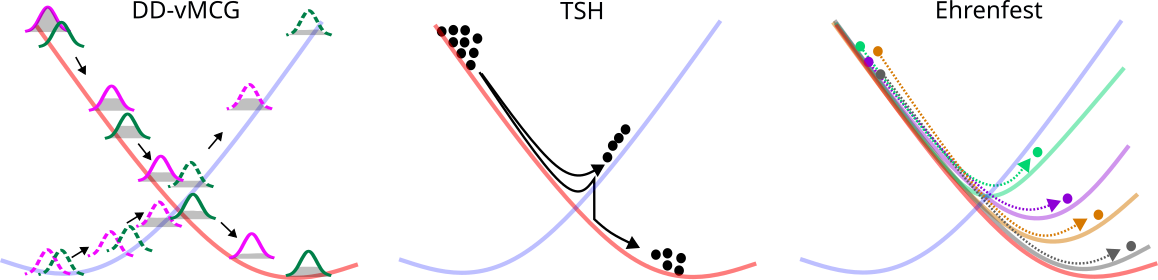}
    \caption{Scheme illustrating the three non-adiabatic dynamics methods employed in this work. (Left) DD-vMCG in the single-set formalism: a set of two interacting Gaussian wavepackets (pink and green) on each state (solid and dashed) are propagated in time and exchange population (grey filling). (Middle) TSH: a swarm of independent trajectories evolves in time following the slope of the ``active'' state curve and can jump from one state to the other. (Right) Ehrenfest: a swarm of independent trajectories evolving each on their \corr{own}\newtxt{individual} time-dependent potential.}
    \label{fig:method}
\end{figure}

Recent works have evaluated the effect of the choice of the dynamics methods and algorithms on femtochemistry with initial excitation to a single electronic state \corr{(and in the absence of quantum nuclear tunneling):}\newtxt{with} little \newtxt{observed} difference \corr{is observed}\newtxt{(in the absence of quantum nuclear tunneling)}.\cite{Ibele2020,Janos2023,Gomez2024}
\newtxt{In the context of attochemistry with initial coherent excitation to a superposition of electronic states, mixed quantum-classical methods, like Ehrenfest and TSH, are commonly employed in the literature.\cite{Meisner-2015,Vacher-2015-JCP,Vacher-2016-FD,Trabattoni-2019,Delgado2021,Arnold2020} However}\corr{In general}, the impact of the approximations underlying \corr{the}\newtxt{such} dynamics methods on simulations of attochemical reactions \corr{is}\newtxt{remains} unknown.
The goal of the present letter is to compare the suitability\corr{for charge-directed reactivity} of several dynamics methods \newtxt{for charge-directed reactivity}: DD-vMCG, TSH, and Ehrenfest (see Figure \ref{fig:method} for a schematic representation). 
While these three methods do not constitute a comprehensive list of all ways to simulate non-adiabatic dynamics available and described in the literature, they cover a diverse range of possible \corr{methods}\newtxt{approaches}. In particular, we have focused on methods that are popular and regularly used within the field of attochemistry; the overarching goal is to evaluate the accuracy of the different dynamics methods specifically within that context.

\begin{figure}[h!]
    \centering
    \includegraphics[width=\textwidth]{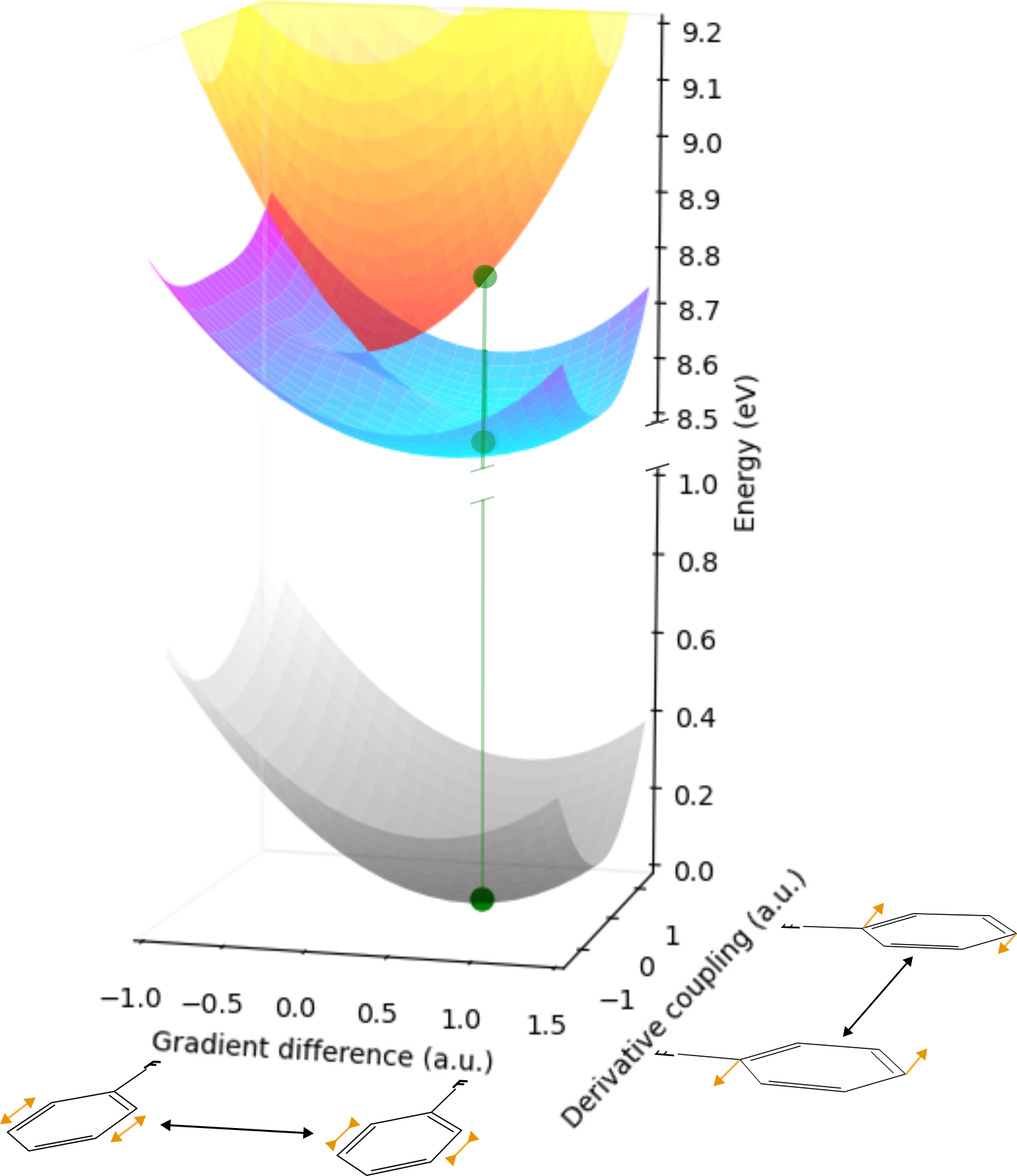}
    \caption{Adiabatic PES of the ground state of the neutral species (grey), $D_0$ (blue-purple) and $D_1$ (red-orange) cationic states along the gradient difference and derivative coupling coordinates with the main in-plane motion depicted on the molecule. The FC geometry is \corr{shown with} \newtxt{indicated by} green dots.}
    \label{fig:PES}
\end{figure}

To assess the performances of the different dynamics methods, this letter focuses on the non-adiabatic dynamics\corr{of fluorobenzene} upon photo-ionization \newtxt{of fluorobenzene}. Attosecond pulses were initially exclusively and are still mainly in the XUV domain, thus leading to molecular ionization.
Fluorobenzene is an ideal system for \newtxt{the} investigation of coupled electron-nuclear dynamics as a conical intersection (CI) between the two lowest-energy cationic states is located in the vicinity of the neutral Franck-Condon (FC) point (see Figure \ref{fig:PES}). 
In general, the passage through a CI can lead to several photo-products\newtxt{,} and the \corr{close proximity}\newtxt{vicinity} of an electronic states\newtxt{'} degeneracy upon ionization here allowed some of the present authors to test the control of nuclear motion near CI via electronic wavepackets.\cite{Ferte_arxiv_2023} The two lowest-energy states result from ionization from the $\pi$ system: the $D_0$ and $D_1$ doublet adiabatic states have a so-called \textit{quinoid} and \textit{anti-quinoid} diabatic characters, respectively, at the FC geometry. \newtxt{The states can also be labeled by their irreducible representation in the $C_{2v}$ point group: $B_1$ for the diabatic \textit{quinoid} state ($\Psi_Q$) and $A_2$ for the diabatic \textit{anti-quinoid} state ($\Psi_A$).} By mapping the PES in the two dimensions spanning the branching plane, derivative coupling \newtxt{($B_2$)} and gradient difference \newtxt{($A_1$)}, the FC point is displaced from the lowest degeneracy point along the \newtxt{totally symmetric motion, i.e., }gradient difference coordinate. At that geometry (see Figure \ref{fig:PES}), the gap between the two cationic states is about 0.27 eV. 

The small size of the chosen system allows us to use a high-accuracy reference dynamics method\newtxt{, namely DD-vMCG,} treating \corr{quantum mechanically}both electron and nuclear dynamics \newtxt{quantum mechanically, along with}\corr{and} accurately \newtxt{capturing} electronic coherence \corr{, namely DD-vMCG,}in full dimensions. 
The results for TSH and Ehrenfest methods shown here in the main manuscript are without decoherence correction and \corr{with inclusion of}\newtxt{include} the derivative coupling in the \newtxt{gradient}\corr{equation-of-motion} for Ehrenfest\corr{(included in the gradient)} and velocity rescaling along the nonadiabatic coupling direction for TSH. While it is standard to simulate photochemical reactions with an ad-hoc decoherence correction, we \corr{think}\newtxt{believe} it may not be appropriate in the present context of attochemistry\newtxt{,} where a superposition of electronic states is initially coherently \corr{excited}\newtxt{populated}. The electronic \corr{populations}\newtxt{coefficients} are propagated in time using explicit calculation of the nonadiabatic coupling vectors for both mixed quantum-classical methods. Results with alternative variants for TSH (for instance\newtxt{,} with a decoherence correction) are shown in SI. 
In the present work, the focus is on the molecular dynamics induced by a coherent electronic superposition; we do not \corr{tackle on}\newtxt{address} the issue of creating the electronic wavepacket with a pulse\newtxt{,} which has been covered from a theoretical point of view in the literature.\cite{Mignolet-2011,Lara-2017,Nisoli-2017}
The dynamics are simulated for 10 fs and are started with two sets of initial electronic wavepackets: a set of pure\corr{diabatic} electronic states either on the \textit{quinoid} ($\Psi_Q$) or \textit{anti-quinoid} ($\Psi_A$) state and a set of mixed \corr{wavepackets}\newtxt{superpositions of them} with a 50-50 weight\corr{s} \corr{with}\newtxt{and} either a positive or negative relative phase: $\frac{1}{\sqrt{2}}(\Psi_Q+\Psi_A)$ and $\frac{1}{\sqrt{2}}(\Psi_Q-\Psi_A)$. \newtxt{It is noted that at the vertical ionization geometry, the chosen diabatic states are equal to the adiabatic ones. Also, we choose here to use the term ``pure'' when referring to a single adiabatic state, and ``mixed'' to a superposition of adiabatic states. Indeed, the adiabatic representation is natural when discussing charge migration and charge-directed reactivity in attochemistry, since these processes only occur when a coherent superposition of \textit{adiabatic} states is populated.}

In the following, we \corr{start}\newtxt{begin} by presenting the set of pure state-induced dynamics\corr{by}\newtxt{,} monitoring the average nuclear motion in the branching space spawned by the two lowest cationic states of fluorobenzene\corr{,} \newtxt{and the electronic population evolution.}\corr{first describing the full quantum results and then discussing those of the mixed quantum-classical approaches.} After\newtxt{ward}, we expand the analysis to dynamics initiated on coherent electronic superpositions and assess the accuracy of the mixed quantum-classical approaches for attochemical applications.

\begin{figure}[h!]
    \centering
    \includegraphics[width=0.9\textwidth]{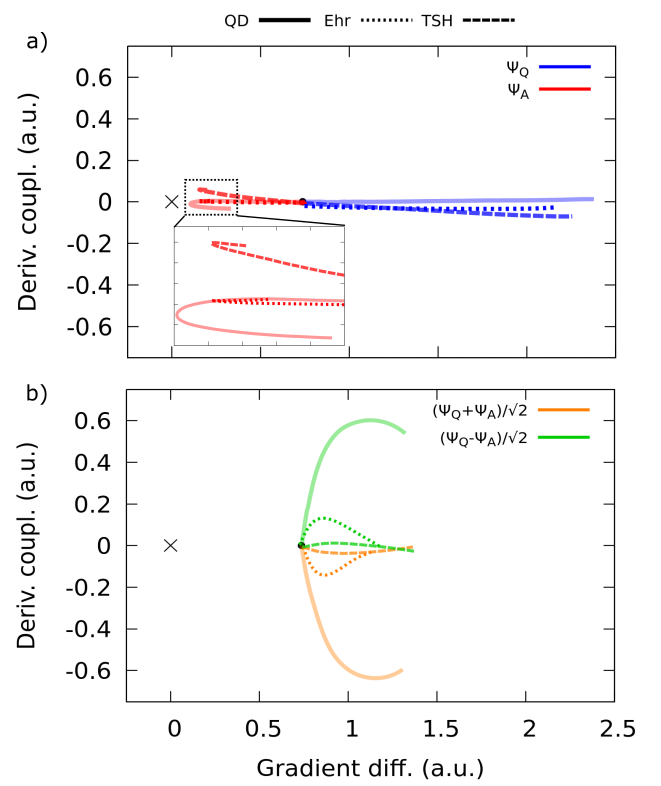}
    \caption{Nuclear motion averaged over all the GWP for DD-vMCG (solid) and trajectories for TSH (dashed) and Ehrenfest (dotted) methods in the branching space coordinates for a set of dynamics initiated (a) on a pure $\Psi_Q$ (blue) or $\Psi_A$ (red) state and (b) on a 50-50 mix of states i.e., $\frac{1}{\sqrt{2}}(\Psi_Q+\Psi_A)$ (orange) or $\frac{1}{\sqrt{2}}(\Psi_Q-\Psi_A)$ (green) state. The CI (cross) is located at the origin and the FC point (dot) \newtxt{is} at $\sim$0.73 along the gradient difference coordinate.}
    \label{fig:BS}
\end{figure}

Figure \ref{fig:BS}a \corr{shows}\newtxt{illustrates} the average nuclear motion in the branching plane \newtxt{as} predicted by the three methods, for the pure cases i.e., exciting \newtxt{to} $\Psi_Q$ (in blue) or $\Psi_A$ (in red)\newtxt{,} \corr{as in}\newtxt{akin to} traditional photochemistry. \newtxt{By initiating the dynamics on an electronic eigenstate, due to symmetry, the center of the nuclear density only evolves along totally symmetric motion ($A_1$ in $C_{2v}$ point group) such as the gradient difference coordinate, until symmetry breaking is allowed by an event such as a passage through a CI. Here, consistently with the shape of the PES, dynamics initiated on the \textit{anti-quinoid} state depicts an initial motion toward the CI along the gradient difference followed by a turning point, while the dynamics on the \textit{quinoid} state move away from the CI. Both cases show negligible motion along the derivative coupling direction. The mixed quantum-classical methods qualitatively reproduce the reference DD-vMCG dynamics results. The main differences are the smaller amplitude of the motion along the gradient difference for the former and the displacement along the derivative coupling at the turning point. In the case of TSH, a slight drift in the direction of the derivative coupling is observed (also across other variants of TSH dynamics, see Figure S4 in SI). While one might initially attribute this drift to the hopping procedure, which rescales the velocity along the derivative coupling in the current procedure to conserve the total energy of the system, a similar trend is observed even when rescaling the full velocity vector. The use of the decoherence correction in TSH shows minimal to negligible impact on the nuclear dynamics within the 10 fs simulation range (see SI).}

\newtxt{The associated adiabatic electronic population evolution is displayed in Figure 4a for the dynamics initiated on $\Psi_A$ (see SI for $\Psi_Q$). In DD-vMCG, the diabatic states are used for the electronic representation, and as such, obtaining the adiabatic state population requires numerical integration as well as renormalization to avoid unphysical populations. The quantum dynamics display a rapid decay of the cationic excited state with the $D_0$ state population dominating past 3.7 fs. The mixed quantum-classical dynamics display a slower decay to the ground state, with Ehrenfest being the slowest. It is noted that at the onset of the dynamics simulation, the adiabatic state populations are not exactly equivalent for the DD-vMCG
wavefunction and classical trajectories; this is attributed to the initial geometries spread of
the latter which is broader than the initial nuclear wavepackets of the quantum dynamics
(see Figures S1 and S2 in SI).}

\begin{figure}[h!]
    \centering
    \includegraphics[width=0.9\textwidth]{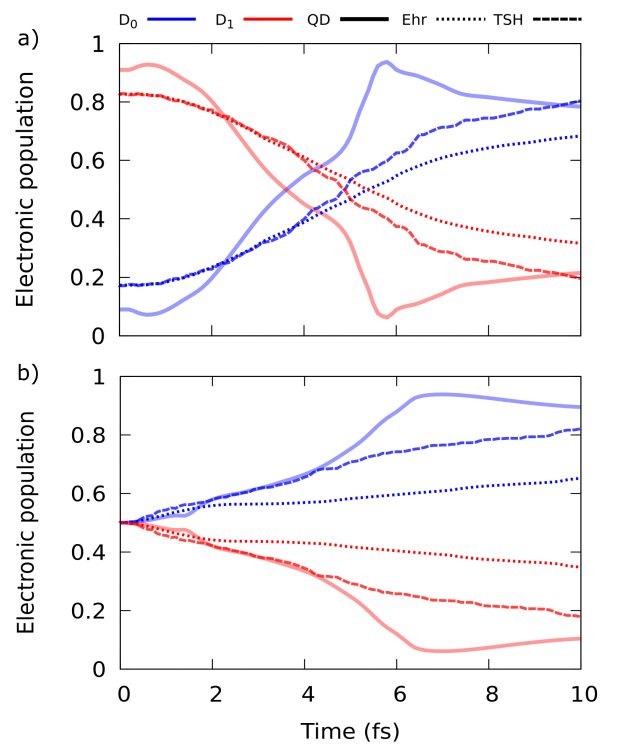}
    \caption{Adiabatic population averaged over all the GWP for DD-vMCG (solid) and trajectories for TSH (dashed) and Ehrenfest (dotted) methods as a function of time for a set of dynamics initiated on (a) the pure $\Psi_A$ state and (b) the 50-50 mix with a positive relative phase $\frac{1}{\sqrt{2}}(\Psi_Q+\Psi_A)$.}
    \label{fig:adia_pop}
\end{figure}

Let us now shift our focus to the dynamics induced by \corr{a }mix\newtxt{ed} electronic wavepacket\newtxt{s}.
Figure \ref{fig:BS}b shows the nuclear dynamics predicted by the three methods following excitation to a coherent superposition of the two lowest cationic states with two different relative phases, representative of an attochemistry experiment. \newtxt{With a coherent superposition of $B_1$ and $A_2$ states (\textit{quinoid} and \textit{anti-quinoid}, respectively), motion along the derivative coupling motion of irreducible character $B_2$ becomes symmetry-allowed and the main driving force of the nuclear dynamics.} \corr{For}\newtxt{Here, in} the reference quantum dynamics (solid curve), a positive relative phase results in a distinct average motion in the negative direction of the derivative coupling\newtxt{,} whereas the superposition with the negative relative phase yields a mirrored dynamics along the positive direction of the derivative coupling. This observation illustrates the concept of charge-directed reactivity sought in attochemistry, i.e., that the nuclear motion can be steered in a specific direction by controlling the composition and phase of the initial electronic wavepacket.

\begin{figure}[h!]
    \centering
    \includegraphics[width=\textwidth]{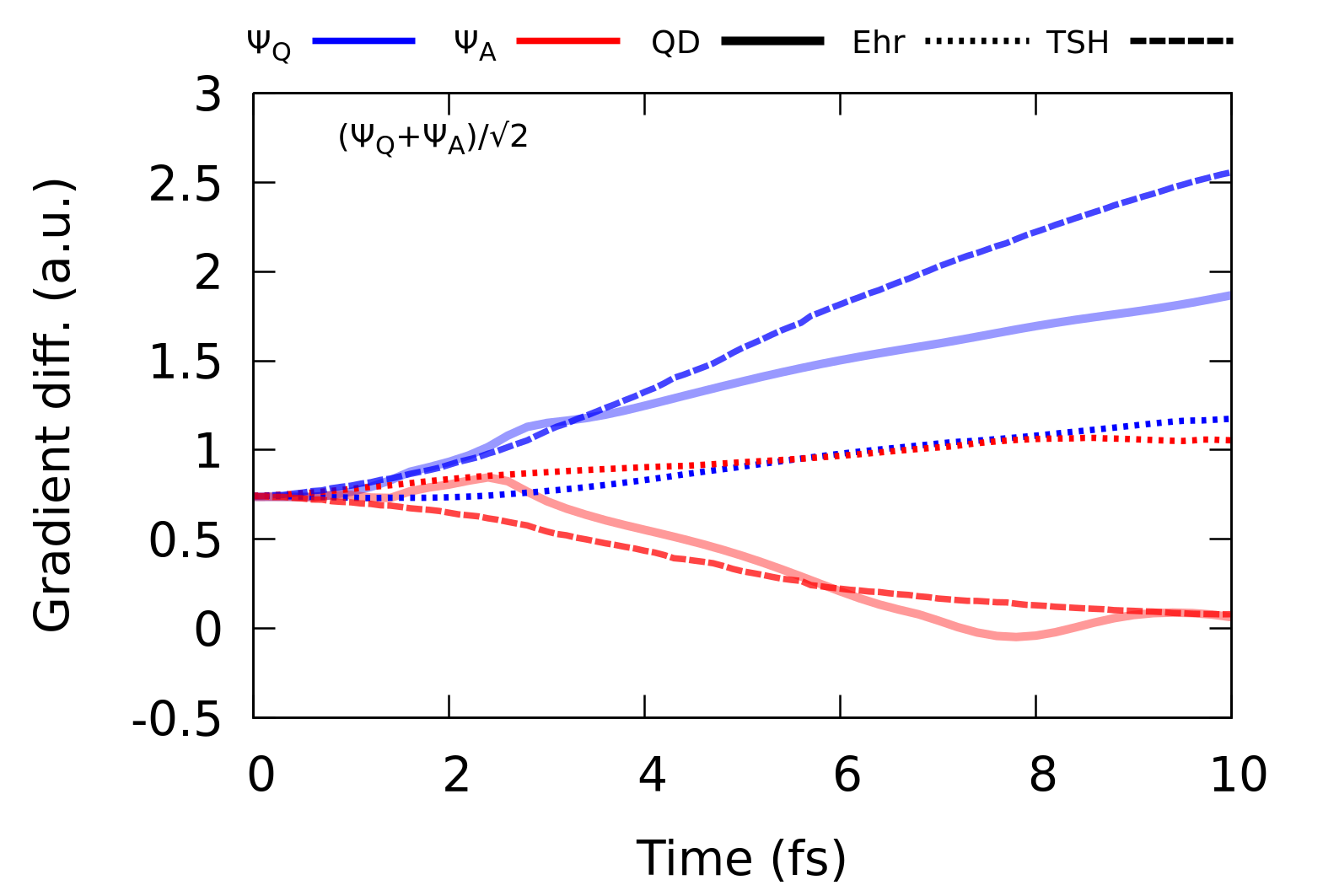}
    \caption{Nuclear motion averaged over all the GWP for DD-vMCG (solid line) and trajectories for TSH (dashed line) and Ehrenfest (dotted line) methods along the gradient difference over time for a set of dynamics induced by $\frac{1}{\sqrt{2}}(\Psi_Q+\Psi_A)$. The average corresponds to a weighted average based on the diabatic states for DD-vMCG and adiabatic states for TSH and Ehrenfest.}
    \label{fig:GD_mix}
\end{figure}

\corr{Along the derivative coupling, the quantum dynamics display an inflection point at an absolute value of approximately 0.6 a.u.~(Figure \ref{fig:BS}b). This can be explained by two reasons. Firstly, there is a development of an adiabatic gradient, coming from the intrastate contribution, pointing toward the minimum. Secondly, the electron dynamics changes the relative phase between the states, resulting in a change to the derivative coupling contribution in both amplitude and direction to the full gradient. 
Both of these effects gradually diminish the amplitude of the nuclear motion and even reverse it along the derivative coupling coordinate, leading to the observed inflection. Additionally, aside from the motions along the derivative coupling, both superpositions of states induce an overall motion in the positive direction of the gradient difference. This is attributed to the facts that the $\Psi_Q$ state induces a motion of larger amplitude than the $\Psi_A$ state (Figure \ref{fig:BS}a) and that the electronic wavepacket gradually de-excites to $\Psi_Q$. In the adiabatic basis, it corresponds to a transfer to the cationic ground state (Figure \ref{fig:adia_pop}b).}

\newtxt{The quantum dynamics exhibits a turning point along the derivative coupling at approximately 0.6 a.u (Figure \ref{fig:BS}b), driven by two factors. Firstly, an adiabatic gradient develops from intrastate contribution, pointing toward the minimum. Secondly, the electron dynamics alters the relative phase between states, modifying the derivative coupling's amplitude and direction. These effects gradually diminish the nuclear motion amplitude, potentially reversing it along the derivative coupling coordinate. Both mixed electronic wavepackets lead to a motion in the positive direction of the gradient difference. This is attributed to $\Psi_Q$ inducing larger motion than $\Psi_A$ (Figure \ref{fig:BS}a), as well as the electronic wavepacket gradually de-exciting to the cationic ground state (Figure \ref{fig:adia_pop}b).  Additionally,}
Figure \ref{fig:GD_mix} \corr{shows}\newtxt{depicts} the average nuclear motion along the gradient difference vector over time induced by $\frac{1}{\sqrt{2}}(\Psi_Q+\Psi_A)$ by projecting the nuclear wavepacket onto a specific electronic state.\corr{The DD-vMCG data correspond to a diabatic weighted average.} The diabatic states generate opposing\corr{driving} forces along the gradient difference coordinate\newtxt{, hence a splitting of the nuclear wavepacket leading to decoherence}. \corr{Consequently, the quantum nuclear density exhibits a distinct separation along this direction, reflecting the character of the electronic states. The nuclear motions being in opposite directions and somewhat similar magnitude along the gradient difference, this could not be seen in the average motion plotted in Figure \ref{fig:BS}b.}

\corr{Both mixed quantum-classical methods display a somewhat similar average nuclear motion along the gradient difference. However, the dynamics along the derivative coupling is qualitatively different (Figure \ref{fig:BS}b). Similarly, the details of the nuclear motion on the different electronic states along the gradient difference differ the most with Ehrenfest (Figure \ref{fig:GD_mix}).
In addition, TSH (dashed curve) and Ehrenfest (dotted curve) predict a non-adiabatic decay of the electronic population to the $D_0$ state but to a smaller extent compared to the reference results (Figure \ref{fig:adia_pop}b); similarly to the pure state induced dynamics case, Ehrenfest shows the slowest electronic population decay.}

\newtxt{We now present the attochemical dynamics predicted by the mixed quantum-classical methods. First, let us explain in more details the initial conditions used to simulate mixed electronic wavepackets. The trajectories are initiated with electronic amplitude coefficients corresponding to the coherent superposition of the two adiabatic states with either a positive or negative relative phase. In addition, in TSH, the trajectories must be initiated on a specific so-called ``active'' state. The choice of an active state is natural when investigating photo-excitation/ionization to specific eigenstates, or multiple states but in an incoherent manner thus simulated individually. Here, for a given initial mixed electronic wavepacket, we have run two sets of TSH simulations, each with a specific adiabatic state as initially active. The TSH results presented here (dashed curve) are the average of the two dynamics sets. 
In general, TSH (dashed curve) and Ehrenfest (dotted curve) exhibit an average nuclear motion along the gradient difference that is similar to the quantum dynamics one (Figure \ref{fig:BS}b).
However, both mixed quantum-classical methods predict qualitatively different nuclear motions along the derivative coupling (Figure \ref{fig:BS}b) and a slower non-adiabatic electronic decay to the $D_0$ state (Figure \ref{fig:adia_pop}) compared to reference results.}

More precisely, the Ehrenfest dynamics show\corr{s} a distinct motion along the derivative coupling, consistent\corr{ly} with the quantum \newtxt{dynamics, albeit with a smaller amplitude.} \corr{results for both initial mix. However, the amplitude of the motion is much smaller.}In comparison, a single FC trajectory with no initial velocity display\newtxt{s} a qualitatively\corr{more} similar motion in the branching space \corr{as}\newtxt{to} a single GWP (see Figure S3 in SI). Thus, with similar initial conditions (geometry, velocity\newtxt{,} and electronic wavepacket), the mixed quantum-classical \newtxt{mean-field} approach\corr{with mean-field} yield\newtxt{s}\corr{a} comparable result\newtxt{s} to quantum dynamics for a single trajectory/GWP.
\corr{The difference in the average for a swarm of trajectories could be explained by the independence of the classical trajectories as well as the Wigner sampling. In the case of the Wigner sampling in phase space, each individual trajectory will experience a different gradient in the derivative coupling direction as well as a different electron dynamics -- the latter can be faster or slower for a larger or smaller electronic state gap, respectively. At the FC point, with a gap of 0.27 eV, the period of the electronic oscillation is about 10 fs. With a larger gap which corresponds to a geometry further away from the CI, a faster electron dynamics is expected which in turn, leads to a faster reversal of the gradient along the derivative coupling direction.}
\newtxt{However, the swarm of}\corr{The} independent \corr{nature of the}trajectories \corr{is}\newtxt{does} not captur\corr{ing}\newtxt{e} the proper quantum feature\corr{of the evolution} of the nuclear wavepacket\newtxt{:}\corr{as these} \newtxt{the classical} trajectories on average show a faster return to the zero value along that coordinate.\corr{with the main contribution coming from the trajectories being far away from the CI. Similar to the pure state-induced dynamics case, the Ehrenfest method show\newtxt{s} the smallest population transfer to the cationic ground state.} 
\newtxt{One driving component of charge-directed reactivity is the adiabatic state energy difference with a larger gap leading to faster electron dynamics, resulting in a quicker reversal of the gradient along the derivative coupling direction. Faster electron dynamics are expected for trajectories further away from the CI.}

\newtxt{Additionally, the individual nuclear motions on different electronic states predicted by Ehrenfest qualitatively differ from the quantum dynamics results (Figure \ref{fig:GD_mix}). It is important to note that the result shown in Figure \ref{fig:GD_mix} for mixed quantum-classical are a weighted average in the adiabatic basis (``active" state for TSH and adiabatic coefficients for Ehrenfest), and thus not directly comparable to the DD-vMCG data in the diabatic basis. However, they provide a good qualitative picture of how the trajectories split based on the main driving electronic state.
The Ehrenfest method is known to provide a long-time inaccurate nuclear density due to its inability to describe bifurcation of the wavepacket through CI and electronic decoherence.\cite{Prezhdo1999,Jasper2006,Akimov2014,Mannouch2023} 
This results here in a lack of splitting along the gradient difference as the trajectories maintain a high degree of mixing in the electronic wavepacket.}

\corr{In Figure \ref{fig:GD_mix}, the Ehrenfest results are based on weighted-average of the adiabatic state coefficient and thus do not compare directly with the DD-vMCG data but give a good qualitative picture on  how the trajectories split based on the main driving electronic state. In the case of Ehrenfest, very little splitting is observed as most of the trajectories keep a fairly  high  mix electronic wavepacket. Thus, the mean-field approach combined with the independent nature of the trajectories prevent the expected splitting of the nuclear dynamics.}

For TSH, \corr{even if it is possible to define a mix initial electronic wavepacket, the trajectories have to be initiated on a specific so-called ``active'' state. The choice of an active state makes sense when investigating photo-excitation/ionization to specific eigenstates, or multiple states but in an incoherent manner so simulated individually. In the present case of an initial coherent electronic wavepacket, we have ran two sets of simulations, each initiated with one adiabatic state as active. The results presented here (dashed curve) are the average of the two dynamics sets. 
The TSH} \newtxt{the} average nuclear motion \newtxt{induced by mixed electronic wavepackets} \corr{shows a similar amplitude along the gradient difference coordinate compared to the full quantum dynamics but it} displays very little to no \corr{-motion}\newtxt{displacement} in the direction of the derivative coupling \newtxt{(Figure \ref{fig:BS}b, dashed curve). The TSH method is thus unable to properly capture the correct effect of the electronic coherence on the nuclear dynamics}. Employing a different velocity rescaling method or decoherence correction yields very similar results (see Figure S4 in SI). 
\corr{The main difference (although still small) is observed upon using a full velocity rescaling while the ad-hoc decoherence correction has very little effect on the dynamics outcome within the 10 fs of simulation.
In} Figure \ref{fig:GD_mix} \corr{, the splitting of the TSH results is done based on the ``active'' state. Although they cannot be compared directly with the DD-vMCG data, they} show\newtxt{s} a \corr{qualitatively similar} split\newtxt{ting}\corr{ted motion} \newtxt{of the nuclear wavepacket} along the gradient difference. \newtxt{The motions are} compar\corr{ed}\newtxt{able} to the reference data \newtxt{although the latter has}\corr{with} a larger amplitude \corr{for dynamics}on the lowest cationic state. \corr{The discrepancy can be attributed to the not perfect equivalence of diabatic to adiabatic state away from the FC point.}
\newtxt{It is noted that} \corr{T}\newtxt{t}he electronic population \newtxt{evolution}\corr{transfer of TSH is very similar to the quantum results for the first 4 fs and shows a smaller decay beyond that point. Very little effect of the mix initial electronic in TSH dynamics is observed on the results compared to the dynamics on} \newtxt{induced upon excitation to mixed electronic wavepackets is similar to the one upon excitation to} pure state\newtxt{s} (see Figure S6 in SI).\corr{A small difference can be seen on the adiabatic population pass the middle point of having 50\% of the trajectories on either state where a smaller decay to the ground state is seen for the TSH dynamics initiated on the \textit{anti-quinoid} state with a mix electronic wavepacket. Using decoherence correction shows a more similar population transfer as their pure state\newtxt{-}induced dynamics equivalent.
For dynamics induced by a coherent superposition of states, the TSH method is unable to properly capture the correct effect of the electronic coherence on the nuclear dynamics independent of the use of decoherence correction. Concerning the electronic population, starting the dynamics with a mix electronic wavepacket yields again a very similar result to the pure state-induced case.} This can be rationali\corr{s}\newtxt{z}ed by how the electron dynamics, i.e., propagation of the equation-of-motions and hopping probabilities are evaluated\newtxt{:}\corr{where} only the predicted change in population \corr{are}\newtxt{is} relevant for hopping procedure in the Tully fewest switch algorithm\corr{ as implemented in SHARC}.\cite{sharc3.0}


In summary, we have tested different dynamics methods to simulate attochemical processes, i.e., non-adiabatic dynamics induced by electronic wavepackets: DD-vMCG, Ehrenfest\newtxt{,} and TSH. We have simulated the coupled electron-nuclear dynamics induced in fluoro-benzene upon ionization and excitation to pure electronic states and \corr{to}coherent mixed \newtxt{electronic wavepackets}\corr{superpositions of them}, taking into account all nuclear dimensions. We have analy\corr{s}\newtxt{z}ed the nuclear motion in the branching space of the nearby CI to investigate the charge-directed reactivity. Mixed quantum-classical methods \corr{are able to}\newtxt{can} reproduce quantitatively the quantum nuclear motion induced by pure electronic \newtxt{a}diabatic states (or an incoherent superposition of them) as in standard photochemistry, consistently with literature.\cite{Ibele2020,Janos2023,Gomez2024} In this case, the resulting nuclear motion is mostly along the gradient difference coordinate, in the positive or negative direction depending on the excited electronic state.

On the other hand, the average nuclear motion induced by a mix\newtxt{ed} electronic wavepacket is characteri\corr{s}\newtxt{z}ed by an initial component mainly along the derivative coupling coordinate (with a splitting of the wavepackets on the different states\corr{ in opposite directions} along the gradient difference coordinate).
This demonstration of charge-directed reactivity in a polyatomic molecule derives from\corr{the} electronic coherence.
The \corr{calculations}\newtxt{most important conclusion} of the present work \corr{show}\newtxt{is} that none of the mixed quantum-classical methods \newtxt{tested here} \corr{is able to}\newtxt{can} capture the full quantum dynamics \corr{of}\newtxt{induced upon} attochemical excitations, in particular the motion along the derivative coupling coordinate\corr{,}\newtxt{. This is} due to the independent nature of the trajectories with initial conditions obtained from a phase space sampling, and propagating only on a single active state in the case of TSH. Moreover, the mean-field approach with Ehrenfest underestimates the adiabatic population transfer to the lowest cationic states\corr{,} and does not describe the splitting of the wavepackets along the gradient difference coordinate. 
Accurate simulation of attochemical dynamics induced by electronic wavepackets thus requires a full quantum treatment to describe properly the effect of electronic coherence on the subsequent nuclear dynamics and the resulting charge-directed reactivity. \corr{The positive conclusion obtained from the current comparative work is}\newtxt{Further, it is interesting to note} that the most accurate dynamics method predicts the strongest attochemical control of the nuclear motion. \newtxt{This was not obvious since mixed quantum-classical methods, suffering from overcoherence, could have overestimated the attochemical control. It is, therefore, a promising result for attochemistry.}

\section{Methods}
The reference results are obtained from a full quantum dynamics of fluorobenzene performed using Direct Dynamics variational Multi-Configurational Gaussian (DD-vMCG) implemented in the QUANTICS package\cite{quantics2020} with the nuclei described by 10 Gaussian wavepackets (GWP) evolving quantum mechanically and the electronic states are diabatised with the regularisation method.
The mixed quantum-classical dynamics is done with the TSH and Ehrenfest methods using the SHARC program\cite{sharc3.0} for a set of 300 trajectories generated from a Wigner sampling. All electronic structure calculations are evaluated with the CASSCF(5e,6o) method with an active space containing the $\pi/\pi^*$ orbitals with a 6-31G* basis set. For DD-vMCG, the electronic structure is evaluated using the Gaussian program\cite{Gaussian16} and for SHARC dynamics, the OpenMolcas program\cite{OpenMolcas-2019} has been employed.
In SHARC, in the absence of an external magnetic field, the doublet states are represented by a pair of degenerate states in either the molecular coulombic Hamiltonian or diagonal basis. Initiali\corr{s}\newtxt{z}ing the dynamics on either one or both degenerate states yield no difference in the present work.
All simulations are done for 10 fs with a nuclear time step of 0.1 fs for TSH and Ehrenfest and adaptative time steps for DD-vMCG using the Runge-Kutta 5$^{th}$ order integrator.

\begin{acknowledgement}
The project is partly funded by the European Union \newtxt{(ERC, 101040356 - ATTOP,} M.V. and T.T.). Views and opinions expressed are however those of the author(s) only and do not necessarily reflect those of the European Union or the European Research Council Executive Agency. Neither the European Union nor the granting authority can be held responsible for them. We also thank the \textit{R\'egion des Pays de la Loire} who provided post-doctoral funding for A.F. The authors thank the CCIPL/Glicid mesocenter installed in Nantes and GENCI-IDRIS (Grant 2021-101353) for \newtxt{the} generous allocation of computational time.
\end{acknowledgement}

\begin{suppinfo}
Trajectories initial conditions, analysis protocol, comparison of single Ehrenfest trajectory and GWP in branching space, result\newtxt{s} for variants of TSH (branching space and electronic adiabatic populations), trajectories densities along gradient difference and derivative couplings as a function of time and convergence plots.
\end{suppinfo}
\bibliography{fbz_tsh_eh_comparison}
\end{document}